\documentclass[twocolumn]{autart} 
\usepackage{graphicx} 
\usepackage{float} 
\usepackage{fancyhdr}
\usepackage{amsfonts}
\usepackage{amstext}
\usepackage{amsmath,amssymb}
\usepackage{booktabs}
\usepackage{url}

\usepackage{wrapfig}
\usepackage[colorlinks=true,linkcolor=blue,citecolor=blue,urlcolor=blue,anchorcolor=blue]{hyperref}
\usepackage[shortlabels]{enumitem}
\setlist[enumerate]{nosep}

\usepackage{natbib}
\setcitestyle{authoryear,round}

\allowdisplaybreaks[4]
\newtheorem{remark}{\rm\textbf{Remark}}
\newtheorem{lemma}{\rm\textbf{Lemma}}
\newtheorem{assumption}{\rm\textbf{Assumption}}
\newtheorem{definition}{\rm\textbf{Definition}}

\newtheorem{theorem}{\rm\textbf{Theorem}}
\newcommand{\tc}{\textcolor{black}}

\newenvironment{proof}{{\noindent\textbf{Proof:}}}{$\hfill{\Box}$}

\begin{document}

\begin{frontmatter}

\title{Coded Event-triggered Control for Nonlinear Systems}

\thanks{This research was supported by the National Research Foundation Singapore under its AI Singapore Programme, Award Number: AISG2-GC-2023-007; Chongqing Top-Notch Young Talents Project under Grant cstc2024ycjh-bgzxm0085; Singapore Maritime Institute (SMI) under its Maritime Transformation Program (MTP) White Space Fund, Project ID: SMI-2022-MTP-08; and also supported by Enabling Future Systems for Offshore Wind Resources (ENFORCE) programme supported by A*STAR under its RIE 2025 Industry Alignment Fund, Grant No: M23M4a0067.
}

\thanks[footnoteinfo]{Corresponding author: Shuzhi Sam Ge.}

\author[Paestum]{Ruihang Ji}\ead{jiruihang@nus.edu.sg},    
\author[Paestum]{Shuzhi Sam Ge}\ead{samge@nus.edu.sg},               
\author[zhaokai]{Kai Zhao}\ead{zhaokai@cqu.edu.cn}  

\address[Paestum]{Electrical and Computer Engineering, National University of Singapore, Singapore 117583, Singapore}             
\address[zhaokai]{School of Automation, Chongqing University, Chongqing 400044, China}


\begin{abstract}                          
This paper studies a Coded Event-triggered Control (CEC) for a class of nonlinear systems under any initial condition. To reduce communication burden, the CEC is designed from the encoding-decoding viewpoint by which only $m$-length string is transmitted for each communication between CEC and actuator. If a more general Entry Capture Problem is encountered, such control design will be rather complicated yet challenging where the performance constraints are satisfied some time after (rather than from the beginning of) system operation, rendering normally employed prescribed performance control invalid because they may be not defined in the initial interval. By introducing auxiliary functions, we develop a Self-adjustable Prescribed Performance (SPP) mechanism which can flexibly adjust the symmetric or asymmetric performance boundaries to accommodate different initial conditions, providing an effective solution for the underlying tracking problem. In this way, the resulted CEC can not only consume less communication resources but also regulate the tracking error under any initial condition into an allowable set before a given time in a bounded and customizable manner. Simulation results verify and clarify the theoretical findings.

\end{abstract}

\end{frontmatter}

\section{Introduction}

Practical systems often operate under limited computational loads but need to maintain certain performance constraints, arising from hardware capability and task requirements, for example, the target tracking problem of aerial robots or the manipulator grasping. Nowadays, event-based control has enticed sustained interest due to its resource efficiency (\cite{wang2022event,aaarzen1999simple,aastrom1999comparison,heemels2012introduction,deng2022distributed,SUN2022110055,liu2023tangent}). As systems grow in complexity, how to save communication resources and guarantee system performance at the same time deserve more in-depth studies.

The event-based control is pioneered in \cite{aaarzen1999simple,aastrom1999comparison}, which emphasizes its advantages over periodic sampling and motivates its systematic design later (\cite{tabuada2007event}). Since then, it has aroused widespread research interest, see, \cite{dimarogonas2011distributed,fan2013distributed,girard2014dynamic,xing2016event,KUMARI2020109163,zhang2022decentralized} and references therein. \tc{However, most existing protocols primarily focus on reducing unnecessary signal transmission but neglect the encoding-decoding process and the associated security concerns for each communication between the control box and the actuator. In practice, when the event condition is met, the real control input is encoded into a lengthy codeword (i.e., 16 bits) before transmission. This could aggravate communication delays and congestion issues due to the limited bandwidth, particularly when communication bits are critical. Moreover, exposing sensitive system signals to public channels raises security concerns.} This motivates us to further study event-triggered control from an encoding-decoding viewpoint, and the following two aspects need to be considered:

\begin{enumerate}[(i)]
\item \textbf{Resource saving}. It might be redundant to transmit such a large number of digits for each communication between control box and actuator, especially when the magnitude of control signal is small. Moreover, in situations where the communication channel can only handle a limited number of bits at a time, lengthy strings for communication can lead to transmission delays and increase vulnerability to packet losses (\cite{mazo2014asynchronous,ding2017overview}). Therefore, it is critical to further save communication resources from the encoding-decoding scheme viewpoint.

\item \textbf{Performance maintenance}. Considering the error induced by the event-trigger scheme, it may degrade the tracking performance. By the prescribed performance control (\cite{bechlioulis2008robust,zhao2017prescribed,ji2022saturation,bu2023fuzzy}, to name a few), the studies of event-triggered control with prescribed performance have been investigated in (\cite{liu2020event,zhang2020event,zhang2022prescribed}). These results are based on an implicit assumption that the prescribed performance should be satisfied from the beginning of system operation. However, in practice, the system's initial conditions often violate the initial performance constraints, rendering these existing methods inapplicable since they are not defined outside the allowable set and the singularity problem is encountered. Therefore, how to develop an event-triggered control with prescribed performance applicable for any initial conditions is still an open problem.

\end{enumerate}

\tc{To handle the first problem, one excellent work in \cite{xing2018event} proposes a 1-bit communication protocol for the relative threshold. Note that this protocol is unable to promptly respond to signal changes due to the high threshold when control inputs are extremely large. As stated in \cite{xing2018event}, the original control signals need to be transmitted sometimes to re-calibrate the decoder due to the signal distortion problem. The other work in \cite{zhang2020fuzzy} studies a 2-bit strategy, switching between fixed and relative thresholds based on the magnitude of control inputs. Therefore, the transmissions of control inputs between control and actuator are necessary, which aggravates the communication burden, especially when the bandwidth is limited. Moreover, all these communication protocols are only applicable to binary systems, limiting their implementation in practice. How to design an effective coded event-triggered scheme for more general base-$p$ number systems to address the above problems is interesting yet challenging.
}

For the second problem, the funnel boundary (\cite{ilchmann2002tracking,berger2022funnel}), the global performance functions (\cite{zhao2021adaptive,chen2020adaptive}), and tuning functions \cite{zhang2021global,ji2023tunnel} can remove such initial limitations. \tc{However, several limitations are observed: (i) The global tracking abilities of funnel boundary and the global prescribed functions are achieved by sacrificing overshoot performance. As their initial performance tends to infinity, it leads to loose performance constraints. (ii) These two methods are only suitable for symmetric performance distribution and rely on specific performance functions, which limits their extensions to more general cases. (iii) Although tuning functions are studied to address a more practical yet challenging Entry Capture Problem, there are no performance constraints on tracking errors during the initial interval. This could lead to potential operational and safety issues. Therefore, how can we regulate any initial tracking error into the allowable set in a bounded and customized manner makes our control design more complicated.
}

In this paper, we propose a CEC and a SPP to reduce the communication resources between control and actuator while addressing the Entry Capture Problem for a class of nonlinear systems. The main contributions lie in:

\begin{enumerate}[(i)]

\item We design a CEC, by which only $m$-length string of base-$p$ number system is transmitted for each communication between control box and actuator. Compared with the existing event-triggered results, the CEC can further reduce the communication burden specifically from the encoding-decoding viewpoint.

\item Different from \cite{zhao2021adaptive,zhang2021global,berger2022funnel}, the proposed SPP can flexibly adjust its performance boundaries in accordance with different initial conditions by introducing auxiliary functions. This feature offers an effective solution to make the control design applicable to any initial conditions, which can be easily extended to other methods.

  \item With the aid of SPP, for any initial condition (including initial-constraint violation), the resulted CEC can handle the Entry Capture Problem for either symmetric or asymmetric performance constraints, by which the tracking error is regulated into an allowable set before a given time in a bounded yet customizable manner rather than no constraints there. In this way, the initial-condition constraints are removed and better transient performance is achieved.

\end{enumerate}

\section{Problem Formulation and Preliminaries}
\subsection{Problem statement}\label{section2.1}

Consider a class of strict-feedback nonlinear systems
\begin{align}\label{Eq1}
\left\{
  \begin{array}{ll}
    \dot x_{i}=f_{i}(\bar x_{i})+g_{i}(\bar x_i)x_{i+1},&i=1,\dots,n-1\\
    \dot x_{n}=f_{n}(\bar x_{n})+g_{n}(\bar x_n)u\\
    y=x_{1}
  \end{array}
\right.
\end{align}
where $x_i\in\mathbb{R}$, $i=1,\dots,n$, is the system state with $\bar x_i=[x_1,\dots,x_i]^T\in\mathbb{R}^i$, $f_i(\cdot):\mathbb{R}^i\rightarrow\mathbb{R}$ is an unknown but continuous function, $g_i(\cdot):\mathbb{R}^i\rightarrow\mathbb{R}$ denotes an unknown time-varying control coefficient, $u\in\mathbb{R}$ and $y\in\mathbb{R}$ are system input and output, respectively.

Define the tracking error $e_1(t) = y(t)-y_d(t)$ with $y_d(t)$ being the desired trajectory. The control objective is to design a CEC for the nonlinear systems \eqref{Eq1} under any initial condition such that:

\begin{enumerate}[(i)]
\item All signals in the closed-loop systems are bounded;

\item Only a coded $m$-length string is required for each communication between CEC and actuator when the event-trigger condition is satisfied; and

\item For any initial condition, the tracking error $e_1(t)$ can be regulated into the prescribed allowable set in a customizable manner before a given time and then be constrained within there.
\end{enumerate}

\begin{definition}\label{definition1}
For any initial tracking condition (including initial constraint violation), if the tracking error $e_1(t)$ is fully constrained by the prescribed performance boundaries right after a user-given settling time $T>0$:
\begin{align}\label{Eq2}
-e_{l}(t)<e_1(t)<e_{u}(t),~\forall t\geq T,
\end{align}
then, such error tracking performance is said to be Entry Capture Problem, where $-e_l(t)$ and $e_u(t)$ denote lower and upper prescribed boundaries, respectively.
\end{definition}

\begin{remark}
The studied Entry Capture Problem represents a frequently encountered tracking situation that the prescribed performance is involved right after a given time $T$, whereas, in most existing works, these performance constraints should be satisfied from the beginning of system operation. One typical example is a flight system which is often released from any condition (including initial-constraint violation), but is required to track and interact with a target in prescribed performance for successful task completion. Therefore, addressing such Entry Capture Problem is interesting yet more challenging since it needs the control design applicable for more general cases.
\end{remark}

\begin{assumption}\label{assumption1}
The desired trajectory $y_d(t)$ and its time derivatives up to $(n+1)$th-order are known, bounded, and piece-wise continuous.
\end{assumption}

\begin{assumption}\label{assumption2}
The nonlinear function $f_i(\bar x_i)$, $i=1,\dots,n$, is unknown but certain crude information is available so that $|f_i(\bar x_i)|\leq b_i\phi_i(\bar x_i)$, $\forall t\geq 0$, where $b_i\geq 0$ is an unknown constant and $\phi_i(\bar x_i) \geq 0$ is a known smooth function.
\end{assumption}

\begin{assumption}\label{assumption3}
The control coefficient $g_i(\bar x_i)$, $i=1,\dots,n$ ,is unknown and time-varying, but away from zero, that is, there exist positive constants $\underline{g}_i$ and $\bar g_i$ such that $0<\underline{g}_i\leq|g_i(\bar x_i)|\leq \bar{g}_i$. Without loss of generality, we assume that the signs of $g_i(\bar x_i)$ are known and all positive.
\end{assumption}

\begin{remark}
Assumption \ref{assumption1} is widely adopted in tracking control of nonlinear systems (\cite{krstic1995nonlinear}). Assumption \ref{assumption2} indicates that some core functions can be easily extracted only based on some crude system information, which is reasonable and in line with practice (\cite{polycarpou1993robust}. Assumption \ref{assumption3} is necessary to guarantee the controllable condition of nonlinear system \eqref{Eq1}, which is made in most control design (\cite{jin2018adaptive}).

\end{remark}

\section{Main Results}
\subsection{Coded Event-triggered Scheme}\label{section31}
With respect to the second objective of CEC, we introduce a Coded Event-triggered Scheme (CES) as follows:
\begin{align}
&u(t) = v(t_k),~\forall t\in [t_k,t_{k+1}),~k=0,1,\dots\label{Eq333}\\
&t_{k+1} = \inf\{t>t_{k} ~|~   |\Delta v(t)|\geq \omega p^\beta  \},\label{Eq444}
\end{align}
where $v(t)$ is the actual control to be developed, $\Delta v(t) = v(t) - u(t)$ denotes the control signal error, $t_k$ represents the update time, $p$ is an even number leading to base-$p$ system (i.e., Binary, Octal number system), the selections of $\beta$ and $\omega$ adopt the following rules:
\begin{align}
&b= q,~~\text{if}~~|u(t)|\in[\pi_q,\pi_{q+1}),\label{khjnds}\\
&\beta =b=\left\{
\begin{array}{ll}
\sum_{j=1}^{m}s_{j,k}p^{j-1}&\text{if}~s_{m,k}<\frac{p}{2},\\
\sum_{j=1}^{m}s_{j,k}p^{j-1}-s_c&\text{if}~s_{m,k}\geq\frac{p}{2},
\end{array}\right.\label{grewdsv}\\
&\omega=\omega_b,\label{iughjvb}
\end{align}
where $q\in\{0,1,\dots,s_c-1\}$, $s_c=\frac{p^m}{2}$ with $m$ being the length of encoded string to be transmitted, as such as there are $s_c$ variations, we specify $0=\pi_0<\pi_1<\dots<\pi_{s_c-1}<\pi_{s_c}=+\infty$ to measure the level of control input, $\omega_0,\dots,\omega_{s_{c}-1}$ are positive constants chosen by designers, and $s_{j,k}\in\{0,1,\dots,p-1\}$ denotes each digit value of base-$p$ number system at $k$-th updated time. \tc{From the triggering condition in \eqref{Eq444}, it becomes essential to determine the sign of $\Delta v(t)$ for accurate encoding-decoding procedures. As shown in the encoding procedure \eqref{grewdsv} and the later decoding procedure \eqref{Eq10}, it indicates that if $\Delta v(t)>0$, let $s_{m,k}<\frac{p}{2}$, otherwise, $s_{m,k}\geq\frac{p}{2}$.} Therefore, supposing $t\in[t_{k},t_{k+1})$, $k=0,1,\dots$, the encoded string $S_{k}$, used for the communication between control and actuator, is constructed by
\begin{align}\label{Eq999}
S_{k}=s_{m,k}s_{m-1,k}\dots s_{2,k}s_{1,k},
\end{align}
where $S_k$ is a $m$-length string to be transmitted whenever the criteria in \eqref{Eq444} is satisfied. \tc{Based on \eqref{Eq444} and \eqref{grewdsv}, $S_k$ can be seen as a sign-and-magnitude representation of the threshold for the control signal error $\Delta v(t)$.} More specifically, $s_{m,k}$ denotes a sign digit, whose value determines the sign of $\Delta v(t)$ (i.e., $\text{if}~s_{m,k}<\frac{p}{2},~\Delta v(t)\geq0$ and $\text{if}~s_{m,k}\geq\frac{p}{2},~\Delta v(t)< 0$); and the magnitude of the threshold for $\Delta v(t)$ is derived by the remaining digital numbers, i.e., $s_{m-1,k},\dots,s_{1,k}$.

\begin{figure}[htbp]
  \centering
  \includegraphics[scale=0.65]{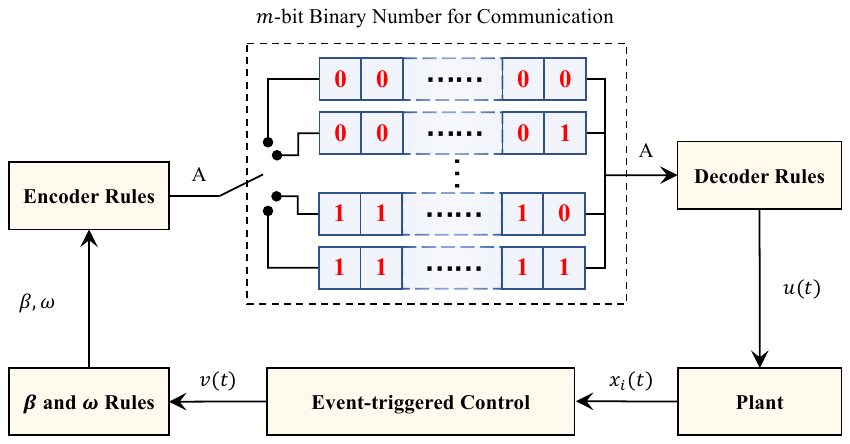}
  \caption{Coded Event-triggered Scheme.}\label{scheme}
\end{figure}

If CES in \eqref{Eq444} is satisfied at $t_{k}$, $S_k$ is expected to be broadcasted to the actuator. When the actuator receives $S_k$, it adopts the following decoder process to update $u_d(t)$ with the aid of the last control input and parameters stored:
\begin{align}\label{Eq10}
u_d(t) = v(t_{k})=\left\{
\begin{array}{ll}
v(t_{k-1})+\omega p^{\beta}&\text{if}~s_{m,k}<\frac{p}{2}\\
v(t_{k-1})-\omega p^{\beta}&\text{if}~s_{m,k}\geq\frac{p}{2}\\
\end{array}\right.
\end{align}
where $u_d(0)=v(t_0)$, and the selections of $\beta$ and $\omega$ follow the same rules in \eqref{khjnds}-\eqref{iughjvb}, which are also stored in the actuator. Based on the triggering condition in \eqref{Eq444} and the decoder rule in \eqref{Eq10}, it can be derived that $u(t)=u_d(t)$. Throughout this paper, we only use $u(t)$ to simplify the control design. In summary, the CES is illustratively described in Fig. \ref{scheme}.

To clearly illustrate the CES, we give a simple case with $p=2$ and $m=3$. This results in a Binary number system and $3$-bit binary string is used for each communication between CEC and actuator, where each digit value of Binary number system has a value of either 1 or 0. The encoding and decoding processes in \eqref{Eq999} and \eqref{Eq10} are executed as shown in Table \ref{ttttable111} whenever the condition in \eqref{Eq444} is triggered.
\begin{table}[htbp]
\centering
\caption{Encoder and decoder design under $p=2$ and $m=3$}
{
\begin{tabular}{l|ll}
\toprule
{Encoder} &\multicolumn{1}{c}{Decoder}\\
\midrule
$S_k=000$  & $u(t)=v(t_k)+\omega p^\beta$,~~$\beta=0$, ~~$\omega=\omega_0$,~~$p=2$\\
$S_k=001$ & $u(t)=v(t_k)+\omega p^\beta$,~~$\beta=1$,~~ $\omega=\omega_1$,~~$p=2$\\
$S_k=010$ & $u(t)=v(t_k)+\omega p^\beta$,~~$\beta=2$, ~~$\omega=\omega_2$,~~$p=2$\\
$S_k=011$ & $u(t)=v(t_k)+\omega p^\beta$,~~$\beta=3$, ~~$\omega=\omega_3$,~~$p=2$\\
$S_k=100$  & $u(t)=v(t_k)-\omega p^\beta$,~~$\beta=0$, ~~$\omega=\omega_0$,~~$p=2$\\
$S_k=101$ & $u(t)=v(t_k)-\omega p^\beta$,~~$\beta=1$, ~~$\omega=\omega_1$,~~$p=2$\\
$S_k=110$ & $u(t)=v(t_k)-\omega p^\beta$,~~$\beta=2$, ~~$\omega=\omega_2$,~~$p=2$\\
$S_k=111$ & $u(t)=v(t_k)-\omega p^\beta$,~~$\beta=3$, ~~$\omega=\omega_3$,~~$p=2$ \\
\bottomrule
\end{tabular}}
\label{ttttable111}
\end{table}

Some salient features of such CES can be observed from the following aspects.
\begin{enumerate}

\item[(i)] \tc{Considering the relative threshold in \cite{xing2018event} with $t_{k+1} = \inf\{t>t_{k}~|~ |\Delta v(t)|\geq \bar \delta|u(t)|+d,0<\bar \delta<1,d>0\}$, it is unable to promptly respond to signal changes due to the high threshold when $|u(t)|$ is extremely large. This signal distortion often degrades tracking performance. A switching threshold scheme is proposed in \cite{xing2016event} which switches between fixed and relative thresholds based on the magnitude of control input. However, it indicates that the transmission of raw control input between control and actuator is necessary, which aggravates the communication burden, especially when the bandwidth is limited. Therefore, our CES is designed from an encoding-decoding viewpoint by which only a concise m-length string is transmitted for each communication instead of the raw control input and $m$ can be chosen by designers. It not only reduces the bandwidth required but also provides a balanced strategy between system performance and network constraints. Please find the following critical point for more details.
}

\item[(ii)] \tc{Different from the fixed threshold in \cite{xing2016event, KUMARI2020109163, zhang2022decentralized} and the above relative threshold where thresholds are either constants or increase monotonically with $|u(t)|$, the proposed CES adopts a piecewise increasing threshold related to $|u(t)|$ and tends to a constant threshold when $|u(t)|$  is excessively large. This dynamic threshold facilitates more accurate tracking performance due to the lower threshold when $|u(t)|$ is small, and ensures a rapid response to the signal's change when $|u(t)|$ is large with the aid of the constant threshold. Therefore, the signal distortion problem commonly observed in fixed and relative thresholds is effectively alleviated since our CES has a good balance between system performance and resource constraints. Comparative simulations are conducted in Section \ref{section6} to illustrate the effectiveness of our CES. Moreover, the fixed threshold can be seen as a special case of our CES if we set $p=2$ and $m=1$ in \eqref{Eq444}. }

\item[(iii)] \tc{The excellent work in \cite{xing2018event} also studies a 1-bit communication protocol for the relative threshold scheme. However, the original control input needs to be transmitted sometimes to re-calibrate the decoder due to the severe signal distortion as the aforementioned discussed. In this paper, our CES not only alleviates the signal distortion problem but also addresses the challenge of co-designing an event-trigger scheme with encoding-decoding rules. As a result, the CES provides a leaner communication protocol, which is friendly for practical systems with limited communication bit resources. On the other hand, by encrypting the control signal into an encoded string $S_{k}$, we inherently improve communication security against potential cyber threats without the knowledge of encoding-decoding rules.}
\end{enumerate}

\subsection{Auxiliary Functions}

To deal with the third control objective of this article, we first introduce the definition of auxiliary functions.

\begin{definition}\label{definition2}
Auxiliary functions $\eta_u(t)$ and $\eta_l(t)$ are scalar functions which satisfies the following properties:
\begin{enumerate}
\item[(i)] $\eta_u^{(k)}$ and $\eta_l^{(k)}$, $k=0,\dots,n+1$, are known, continuous, and bounded;
\item[(ii)] $\eta_u(0)>e_1(0)-e_u(0)$ and $\eta_l(0)<e_1(0)+e_l(0)$;
\item[(iii)] $\eta_u(t)$, $\eta_l(t)$, $\dot\eta_u(t)$, $\dot\eta_l(t)\rightarrow0$ as $t\rightarrow T$, where $T$ is the settling time in \eqref{Eq2}; and 
\item[(iv)] $\eta_u(t)\geq\eta_l(t)$, $\forall t\geq 0$ and $\eta_u(t)=\eta_l(t)=0$, $\forall t\geq T$.
\end{enumerate}
\end{definition}

Obviously, there exist many candidates satisfying these properties, for example,
\begin{align}
&\eta_u(t)=\left\{
\begin{array}{ll}
(e_1(0)+\frac{(\lambda-2)e_u(0)+\lambda e_l(0)}{2})e^{-\frac{l Tt}{T-t}},&0\leq t<T\\
0,&t\geq T
\end{array}\right.\notag\\
&\eta_l(t)=\left\{
\begin{array}{ll}
(e_1(0)-\frac{\lambda e_u(0)+(\lambda-2) e_l(0)}{2})e^{-\frac{l Tt}{T-t}},&0\leq t<T\\
0,&t\geq T
\end{array}\right.\notag
\end{align}
where $l$ and $\lambda\geq1$ are positive constants, $e_l(0)$ and $e_u(0)$ denote the initial values of performance constraints $e_l(t)$ and $e_u(t)$ which will be defined in the next section. Throughout this paper, we use the above functions as the auxiliary functions.

\begin{remark}
More details about the motivation of such auxiliary functions are provided here. From the third objective in section \ref{section2.1}, the control scheme should be adapted to any initial condition. However, in practice, the initial condition may violate the performance constraints initially, rendering most existing PPC methods inapplicable since they are not defined and suffer from singularity problem in such scenario. To handle this problem, a straightforward approach is to utilize these auxiliary functions to adjust the performance boundaries in accordance with the initial condition. By leveraging the second property in Definition \ref{definition2}, any given initial condition would remain within an updated and allowable set, thereby ensuring the applicability of the control method. Other properties are also crucial for system analysis under addressing the Entry Capture Problem \eqref{Eq2}. Moreover, there would be fruitful expressions of such additional auxiliary functions which facilitate their extension to other control approaches.
\end{remark}

\subsection{Self-adjustable Prescribed Performance}
We design the following Self-adjustable Prescribed Performance (SPP) on the tracking error $e_1(t)$ by introducing the above auxiliary functions into the performance boundaries:
\begin{align}\label{Eq5}
-E_l(t)<e_1(t)<E_u(t),~\forall t\geq 0,
\end{align}
with
\begin{align}
&E_u(t) = e_u(t)+\eta_u(t),\label{Eq6}\\
&E_l(t) = e_l(t)-\eta_l(t),\label{Eq7}
\end{align}
where $e_u(t)$ and $-e_l(t)$ represent the original upper and lower Tunnel Prescribed Performance (TPP) (\cite{ji2022saturation2}), which are given by:
\begin{align}
&e_{u}(t)=(\delta+\text{sign}(e_{1,0}))\rho(t)-\rho_{\infty}\text{sign}(e_{1,0}),\label{Eq8}\\
&e_{l}(t)=(\delta-\text{sign}(e_{1,0}))\rho(t)+\rho_{\infty}\text{sign}(e_{1,0}),\label{Eq9}
\end{align}
where $0<\delta<1$, $e_{1,0}=e_1(0)$, $\rho(t)=(\rho_{0}-\rho_{\infty})e^{-\varsigma t}+\rho_{\infty}$ with $\rho_0>\rho_\infty>0$ and $\varsigma>0$.

\begin{figure*}[htbp]
  \centering
  \includegraphics[scale=0.75]{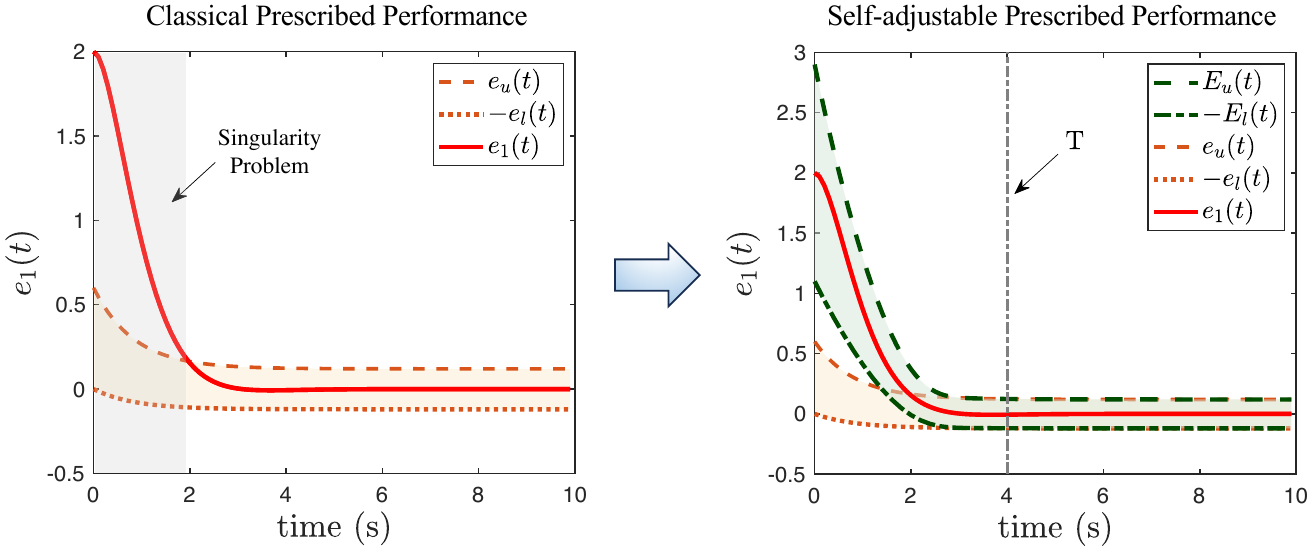}
  \caption{The mechanism behind the Self-adjustable Prescribed Performance.}\label{SPP}
\end{figure*}

To better illustrate the mechanism behind such SPP, we take $e_1(0)>0$ as an example. The parameters of SPP \eqref{Eq6}-\eqref{Eq7} are selected as: $\delta=0.6$, $\rho_0=0.5$, $\rho_\infty=0.2$, $l=1$, $\varsigma=1.2$, $\lambda=3$, $T=4$, and the initial tracking condition is $e_1(0) = 2$. As shown in the left plot of Fig. \ref{SPP}, the initial tracking error $e_1(0)$ violates the original allowable set established by $e_u(t)$ and $-e_l(t)$, which is colored in brown. It renders the previous methods inapplicable due to the singularity problem encountered. With the aid of the auxiliary functions, SPP is capable of re-adjusting the performance boundaries according to different initial conditions such that $e_1(t)$ is always within the updated allowable set initially (i.e., $-E_l(0)<e_1(0)<E_u(0)$) as shown in the right plot of Fig. \ref{SPP} in green color. During the initial time interval ($0\leq t\leq T$), our SPP not only ensures the control method applicable for any initial conditions but also provides temporary performance constraints on the tracking error. When $t\geq T$, SPP is equivalent to the original performance boundaries, that is, $E_u(t)=e_u(t)$ and $-E_l(t)=-e_l(t)$ since $\eta_u(t)=\eta_l(t)=0$ as depicted in the right figure. Therefore, SPP provides an effective solution for the underlying Entry Capture Problem \eqref{Eq2} as stated in the following lemma.

\begin{lemma}\label{lemma1}
If $-E_l(t)<e_1(t)<E_u(t)$ holds for all $t\geq 0$, then the Entry Capture Problem \eqref{Eq2} is obtained.
\end{lemma}
 
\begin{proof}
When $0\leq t<T$, the boundedness of $\eta_u(t)$ and $\eta_l(t)$ is guaranteed according to Definition \ref{definition2}. By \eqref{Eq8}-\eqref{Eq9}, the original TPP is also bounded there. It ensures that $e_1(t)$ is bounded on $[0,T)$. When $t\geq T$, we have $-E_l(t)<e_1(t)<E_u(t)$ which can be rewritten as $-e_l(t)<e_1(t)<e_u(t)$. If the above statement holds, the Entry Capture Problem is, therefore, addressed by invoking Definition \ref{definition1}.
\end{proof}

From the above discussion, the control objective is successfully transfered to guarantee $e_1(t)$ evolving within the SPP envelope \eqref{Eq5} to deal with the issue of Entry Capture Problem.

\begin{remark}
In comparison with funnel boundary (\cite{berger2022funnel}), global prescribed performance (\cite{zhao2021adaptive,zhang2021global}) and tuning functions (\cite{zhang2021global,ji2023tunnel}), some salient features of the proposed SPP \eqref{Eq5} are observed as follows:
\begin{enumerate}[(i)]

\item \textbf{(Improved transient performance)} The funnel boundary and global prescribed performance often encounter an overshoot problem, as their initial performance constraints tend to infinity leading to a loose allowable set. Similarly, by tuning functions, there are no constraints on tracking errors during the initial interval, which is undesirable in practice due to safety concerns and task demands. \tc{Different from these methods, our SPP is capable of adjusting performance boundaries with the help of auxiliary functions such that the proposed control scheme can be applied to any initial condition even if the initial constraints are violated. It provides bounded yet customizable virtual performance boundaries during the initial interval as depicted in Fig. \ref{SPP}, effectively avoiding the aforementioned severe overshoot performance and unconstrained behaviors. The tracking error is regulated into the allowable set within finite time and Entry Capture Problem is therefore solved.}

\item \textbf{(Flexibility and Extendibility)} \tc{Our SPP allows for an easy extension to other control methods by integrating the introduced auxiliary functions into the performance boundaries as formulated in \eqref{Eq5}-\eqref{Eq7}. However, the funnel boundary and global prescribed performance rely on specific performance functions, which limits their adaptability to other control methods. Moreover, the previous methods are limited to symmetric performance distributions but the proposed SPP is applicable to both symmetric and asymmetric cases, thereby enhancing its applicability to more general cases. In this paper, we employ asymmetric TPP as the baseline, which not only provides a tighter allowable set but also limits overshoot performance during the initial interval of the Entry Capture Problem.
}
\end{enumerate}
\end{remark}

\section{Coded Event-triggered Control for Second-order System}

To clearly illustrate our design methodology, we first consider the following second-order nonlinear system:
\begin{align}
\left\{
  \begin{array}{ll}
    \dot x_{1}=f_{1}(\bar x_{1})+g_{1}(\bar x_1)x_{2},\\
    \dot x_{2}=f_{2}(\bar x_{2})+g_{2}(\bar x_2)u,\label{Eq1155}
  \end{array}
\right.
\end{align}
where $x_1$ and $x_2$ are system states, $f_i(\bar x_i)$ and $g_i(\bar x_i)$, $i=1,2$, are unknown yet smooth nonlinear functions satisfying Assumptions \ref{assumption2}-\ref{assumption3}, and $u$ denotes the system control input.

\textbf{Step 1:} We first define the tracking errors:
\begin{align}
&e_1=x_1-y_d\label{Eq1166}\\
&e_2=x_2-\alpha_1,\label{Eq17}
\end{align}
where $\alpha_1$ represents the virtual control input to be defined shortly. In order to guarantee $e_1$ satisfying SPP \eqref{Eq5} for all $t\geq 0$, we introduce the following error transformation function
\begin{align}\label{Eq1188}
z_1=\ln \left( \frac{E_l+e_1}{E_u-e_1}\right)
\end{align}
then, by invoking \eqref{Eq6}-\eqref{Eq7} and \eqref{Eq17}, its time derivative is
\begin{align}\label{Eq19}
\hspace{-0.1cm}\dot z_1&=\mu_1\dot e_1+\mu_2=\mu_1(f_1+g_1e_2-\dot y_d+g_1\alpha_1)+\mu_2
\end{align}
where $\mu_1=\ell(E_u+E_l)$ and $\mu_2=\ell((\dot e_l-\dot\eta_l)(E_u-e_1)-(\dot e_u+\dot\eta_u)(E_l+e_1))$ and $\ell = \frac{1}{(e_1+E_l)(E_u-e_1)}$. From \eqref{Eq6}-\eqref{Eq9} and the property (iv) in Definition \ref{definition2}, it can be derived that $E_u(t)+E_l(t)>0$ and $(E_u(t)-e_1(t))(E_l(t)+e_l(t))>0$ are bounded functions for $e_1(t)$ in the compact set $\Omega_{e_1}=\{e_1(t)\in\mathbb{R}:-E_l(t)<e_1(t)<E_u(t)\}$. Therefore, we have $\mu_1\neq0$ and $\mu_1\in L_\infty$. Then, the time derivative of $\frac{1}{2}z_1^2$ along \eqref{Eq19} is:
\begin{align}\label{Eq20}
z_1\dot z_1&=z_1\mu_1(f_1+g_1e_2-\dot y_d)+z_1\mu_1g_1\alpha_1+z_1u_2\notag\\
&=z_1\mu_1g_1\alpha_1+\Xi_1,
\end{align}
where $\Xi_1=z_1\mu_1(f_1+g_1e_2-\dot y_d)+z_1\mu_2$. Upon using Assumptions \ref{assumption1}-\ref{assumption3} and Young's inequality, we have
\begin{align}
&-z_1\mu_1\dot y_d\leq \underline{g}_1z_1^2\mu_1^2\dot y_d^2+ \frac{1}{4\underline{g}_1},\label{Eq2211}\\
&z_1\mu_{1}f_1\leq \underline{g}_1z_1^2\mu_{1}^2b_1^2\phi_1^2+\frac{1}{4\underline{g}_1},\\
&z_1\mu_{1}g_1e_2\leq \underline{g}_2z_1^2\mu_{1}^2e_2^2+\frac{\bar g_1^2}{4\underline{g}_2},\\
&z_1\mu_2\leq \underline{g}_1z_1^2\mu_2^2+\frac{1}{4\underline{g}_1}.\label{Eq2244}
\end{align}
Therefore, $\Xi_1$ in \eqref{Eq20} is bounded by
\begin{align}\label{Eq25}
\Xi_1\leq \underline{g}_1\theta_1 z_1^2\Phi_1+\underline{g}_2z_1^2\mu_{1}^2e_2^2+ \frac{3}{4\underline{g}_1}+\frac{\bar g_1^2}{4\underline{g}_2}
\end{align}
with
\begin{align}
&\theta_1=\max\left\{1,b_1^2\right\},\label{eq44554455}\\
&\Phi_1=\mu_1^2\dot y_d^2+\mu_1^2\phi_1^2+\mu_2^2.
\end{align}

We develop the virtual control input $\alpha_1$ as
\begin{align}
	&\alpha_1=-\frac{1}{\mu_1}(c_1z_1+z_1\hat\theta_1\Phi_1),\label{Eq28}\\
	&\dot{\hat\theta}_1 = r_1z_1^2\Phi_1-\sigma_1\hat\theta_1,~~\hat\theta_1(0)\ge 0 \label{Eq29}
\end{align}
where $c_1$, $r_1$, and $\sigma_1$ are positive constants, and $\hat\theta_1$ is the estimation of $\theta_1$.

Consider the following Lyapunov function
\begin{align}\notag
V_1=\frac{1}{2}z_1^2+\frac{\underline{g}_1}{2r_1}\tilde\theta_1^2,
\end{align}
where $\tilde\theta_1=\theta_1-\hat\theta_1$. Then, from \eqref{Eq20}, \eqref{Eq25}, \eqref{Eq28} and \eqref{Eq29}, its time derivative can be derived as
\begin{align}
\hspace{-0.2cm}\dot V_1 =& z_1\mu_1g_1\alpha_1+\Xi_1-\frac{\underline{g}_1}{r_1}\tilde\theta_1\dot{\hat\theta}_1\notag\\
\leq& -c_1\underline{g}_1z_1^2+\underline{g}_2z_1^2\mu_{1}^2e_2^2+ \frac{3}{4\underline{g}_1}+\frac{\bar g_1^2}{4\underline{g}_2}+\frac{\underline{g}_1\sigma_1}{r_1}\tilde\theta_1\hat\theta_1.\notag
\end{align}
Based on the definition of $\tilde\theta_1$, it yields
\begin{align}
&\tilde\theta_1\hat\theta_1=\tilde\theta_1(\theta_1-\tilde\theta_1)\leq -\frac{1}{2}\tilde\theta_1^2+\frac{1}{2}\theta_1^2.\notag
\end{align}
We then have
\begin{align}\label{Eq33}
\dot V_1\leq& -c_1\underline{g}_1z_1^2-\bar\sigma_1\tilde\theta_1^2+\underline{g}_2z_1^2\mu_{1}^2e_2^2+\varepsilon_1,
\end{align}
where $\bar\sigma_1=\frac{\underline{g}_1\sigma_1}{2r_1}$ and $\varepsilon_1=\frac{3}{4\underline{g}_1}+\frac{\bar g_1^2}{4\underline{g}_2}+\frac{\underline{g}_1\sigma_i}{2r_1}\theta_1^2$ is a bounded signal. Notice that the item $\underline{g}_2z_1^2\mu_{1}^2e_2^2$ will be tackled in the next step.

\textbf{Step 2:} From \eqref{Eq1155} and \eqref{Eq17}, the time derivative of $e_2$ is
\begin{align}
\dot e_2=\dot x_2-\dot\alpha_1=f_2+g_2u-\dot\alpha_1,\label{Eq34}
\end{align}
with
\begin{align}\label{Eq35}
\dot\alpha_1=\frac{\partial\alpha_1}{\partial x_1}(f_1+g_1x_2)+\Delta\alpha_1,
\end{align}
where $\Delta\alpha_1=\sum_{k=0}^1\frac{\partial\alpha_1}{\partial y_d^{(k)}}y_d^{(k+1)} +\sum_{k=0}^1\frac{\partial\alpha_1}{\partial \rho^{(k)}}\rho^{(k+1)}+\sum_{k=0}^1\frac{\partial\alpha_1}{\partial \eta_u^{(k)}}\dot \eta_u^{(k+1)}+\sum_{k=0}^1\frac{\partial\alpha_1}{\partial \eta_l^{(k)}}\dot \eta_l^{(k+1)}+\frac{\partial\alpha_1}{\partial\hat\theta_1}\dot{\hat\theta}_1$, which is computable. According to the definition of $\Delta v(t)$, we can obtain
\begin{align}\label{Eq36}
u(t)=v(t)-\Delta v(t),~|\Delta v(t)|\leq \bar p,
\end{align}
where $\bar p=\max\{\omega_0 p^0,\omega_1 p,\dots,\omega_{s_c} p^{s_c} \}$ is a positive constant. We then consider the Lyapunov function $V_2$:
\begin{align}
V_2= V_1+\frac{1}{2}e_2^2+\frac{\underline{g}_2}{2r_2}\tilde\theta_2^2,
\end{align}
where $\tilde\theta_2=\theta_2-\hat\theta_2$, $\hat\theta_2$ is the estimation of $\theta_2$ in \eqref{Eq45}, and $r_2$ is a positive constant. The time derivative of $V_2$, along \eqref{Eq34}-\eqref{Eq36}, is derived
\begin{align}\label{Eq38}
\hspace{-0.2cm}\dot V_2=&\dot V_1+e_2\dot e_2-\frac{\underline{g}_2}{r_2}\tilde\theta_2\dot{\hat\theta}_2\notag\\
\leq&-c_1\underline{g}_1z_1^2-\bar\sigma_1\tilde\theta_1^2+e_2g_2v-\frac{\underline{g}_2}{r_2}\tilde\theta_2\dot{\hat\theta}_2+\Xi_2+\varepsilon_1
\end{align}
where $\Xi_2=e_2(f_2-\frac{\partial\alpha_1}{\partial x_1}(f_1+g_1x_2)-\Delta\alpha_1)-e_2g_2\Delta v+\underline{g}_2z_1^2\mu_{1}^2e_2^2$. Using Young's inequality, we can also expand $\Xi_2$ as inequalities in \eqref{Eq2211}-\eqref{Eq2244}. Note that the control input signal will be updated whenever the coded event-triggered scheme \eqref{Eq444} is triggered, which indicates that $|\Delta v|\leq\omega p^\beta$ holds. As $\omega$, $\beta$ and $p$ are all bounded numbers, from \eqref{Eq36}, it can be derived that $\bar p$ is a bounded constant, leading to:
\begin{align}
&-e_2g_2\Delta v\leq \underline{g}_2\bar e_2^2+\frac{\bar g_2^2}{4\underline{g}_2}\bar p^2.
\end{align}
It yields that
\begin{align}\label{Eq44}
\Xi_2\leq \underline{g}_2\theta_2e_2^2\Phi_2+\frac{3}{4\underline{g}_2}+\frac{\bar g_1^2}{4\underline{g}_2}+\frac{\bar g_2^2}{4\underline{g}_2}\bar p^2,
\end{align}
where 
\begin{align}
&\hspace{-0.25cm}\theta_2=\max\left\{1,b_1^2,b_2^2\right\},\label{Eq45}\\
&\hspace{-0.25cm}\Phi_2=(\frac{\partial\alpha_{1}}{\partial x_1}\phi_1)^2+(\frac{\partial\alpha_{1}}{\partial x_1}x_{2})^2\!+\!(\Delta\alpha_{1})^2+\mu_1^2z_{1}^2\!+\!\phi_2^2\!+\!1.
\end{align}

To move forward, the actual control is developed as
\begin{align}
&v(t) =
-(c_2e_2+\hat\theta_2e_2\Phi_2),
\label{Eq47}
\\
&\dot{\hat\theta}_2 = r_2e_2^2\Phi_2-\sigma_2\hat\theta_2,~~\hat\theta_2(0)\ge 0\label{Eq48}
\end{align}
where $c_2,r_2,\sigma_2>0$ and $\hat\theta_2$ is the estimation of $\theta_2$. By \eqref{Eq44}-\eqref{Eq48}, the inequality \eqref{Eq38} can be rewritten as
\begin{align}\label{Eq49}
\dot V_2
\leq&-c_1\underline{g}_1z_1^2-\bar\sigma_1\tilde\theta_1^2-\underline{g}_2c_2e_2^2+\frac{\sigma_2\underline{g}_2}{r_2}\tilde\theta_2\hat\theta_2\notag\\
&+\frac{3}{4\underline{g}_2}+\frac{\bar g_1^2}{4\underline{g}_2}+\frac{\bar g_2^2}{4\underline{g}_2}\bar p^2+\varepsilon_1.
\end{align}
Based on the definition of $\tilde\theta_2$, we obtain
\begin{align}\label{Eq50}
&\tilde\theta_2\hat\theta_2=\tilde\theta_2(\theta_2-\tilde\theta_2)\leq -\frac{1}{2}\tilde\theta_2^2+\frac{1}{2}\theta_2^2.
\end{align}
By substituting \eqref{Eq50} into \eqref{Eq49} yields
\begin{align}\label{Eq51}
\dot V_2
\leq&-c_1\underline{g}_1z_1^2-\bar\sigma_1\tilde\theta_1^2-\underline{g}_2c_2e_2^2-\bar\sigma_2\tilde\theta_2^2+\varepsilon_2,
\end{align}
where $\bar\sigma_2=\frac{\underline{g}_2\sigma_2}{2r_2}$ and $\varepsilon_2=\frac{\underline{g}_2\sigma_2}{2r_2}\theta_2^2+\frac{3}{4\underline{g}_2}+\frac{\bar g_1^2}{4\underline{g}_2}+\frac{\bar g_2^2}{4\underline{g}_2}\bar p^2+\varepsilon_1$.

To summarize, we establish the following theorem.

\begin{theorem}\label{theorem1}

Consider the second-order nonlinear system \eqref{Eq1155} under Assumptions \ref{assumption1}-\ref{assumption3}. The virtual control $\alpha_1$ and the actual control input $v(t)$ are developed in \eqref{Eq28} and \eqref{Eq47}. The adaptive laws are given by \eqref{Eq29} and \eqref{Eq48}. Following the Coded Event-triggered Scheme \eqref{Eq333}-\eqref{Eq444}, the proposed control method guarantees the following.
\begin{enumerate}[(i)]

\item The tracking error $e_1$ satisfies Entry Capture Problem property since it is strictly constrained by the prescribed performance: $-e_{l}(t)<e_1(t)<e_{u}(t)$ right after an user-given finite time $T$ as presented in Definition \ref{definition1};

\item All signals in the closed-loop system are guaranteed to be bounded regardless of initial conditions; and

\item Zeno behavior is excluded.
\end{enumerate}
\end{theorem}

\begin{proof}
By revisiting \eqref{Eq51}, there is
\begin{align}
\dot V_2
\leq&-c_1\underline{g}_1z_1^2-\bar\sigma_1\tilde\theta_1^2-\underline{g}_2c_2e_2^2-\bar\sigma_2\tilde\theta_2^2+\varepsilon_2\notag\\
\leq&-\pi V_2+\varepsilon_2,\notag
\end{align}
where $\pi=\min\{2c_1\underline{g}_1,2c_2\underline{g}_2,2\frac{r_1\bar\sigma_1}{\underline{g}_1},2\frac{r_2\bar\sigma_2}{\underline{g}_2}\}>0$. Subsequently, we can have
\begin{align}
0\leq V_2(t)\leq \frac{\varepsilon_2}{\pi}+(V_2(0)-\frac{\varepsilon_2}{\pi})e^{-\pi t}.\notag
\end{align}
It can be concluded that $V_2$ is bounded which guarantees the boundedness of $z_1$, $e_2$, $\tilde\theta_1$, and $\tilde\theta_2$. From the error transformation function in \eqref{Eq1188}, it is certain that $-E_l(t)<e_1(t)<E_u(t)$ holds due to $z_1\in L_{\infty}$. According to Lemma \ref{lemma1}, $e_1$ satisfies $-e_l(t)<e_1(t)<e_u(t)$ for $t\geq T$, which indicates that $e_1$ follows the Entry Capture Problem in Definition \ref{definition1}.

Next, other tracking signals in the closed-loop system are proved to be bounded. Considering the adaptive laws \eqref{Eq29} and \eqref{Eq48}, $\theta_1,\theta_2,\dot{\hat\theta}_1$ and $\dot{\hat\theta}_2$ remain bounded. The virtual control input $\alpha_1$ defined in \eqref{Eq28} belong to $L_\infty$. Moreover, by \eqref{Eq47}, the boundedness of actual control signal $v(t)$ is guaranteed. From \eqref{Eq1166} and \eqref{Eq17}, it can be derived that $x_1$ and $x_2$ also keep bounded. All the closed-loop signals are therefore bounded for the second-order nonlinear system under any initial condition.

To illustrate the exclusion of Zeno behavior, the time derivative of $\Delta v(t)$ during the inter-execution interval is given
\begin{align}
 \Delta \dot v(t)=\dot v(t)-\dot v(t_k)=\dot v(t),~t_k\leq t<t_{k+1},\notag
\end{align}
where $k=1,2,\dots$ By \eqref{Eq47}, it yields $\dot v(t)=\sum_{k=1}^{2}\frac{\partial v}{\partial x_k}\dot x_k+\sum_{k=0}^{2}\frac{\partial v}{\partial y_d^{(k)}}y_d^{(k+1)}+\sum_{k=1}^{2}\frac{\partial v}{\partial\hat\theta_k}\dot{\hat\theta}_k +\sum_{k=0}^{2}\frac{\partial v}{\partial \rho^{(k)}}\rho^{(k+1)}+\sum_{k=0}^2\frac{\partial v}{\partial \eta_u^{(k)}}\dot \eta_u^{(k+1)}+\sum_{k=0}^2\frac{\partial v}{\partial \eta_l^{(k)}}\dot \eta_l^{(k+1)}$. In accordance with Assumptions \ref{assumption1}-\ref{assumption3}, Definition \ref{definition2}, prescribed performance functions $e_u(t)$ and $e_l(t)$ defined in \eqref{Eq8}-\eqref{Eq9}, and the fact that $\dot e_2$ governed by \eqref{Eq34} and other closed-loop signals remain bounded, the boundedness of $ \dot v$ is ensured. For convenience, the upper bound of $| \dot v|$ can be specified by a positive constant $\kappa$. Therefore, we can obtain
\begin{align}
|\Delta \dot v(t)|\leq \kappa,~t_k\leq t<t_{k+1},\notag
\end{align}
where $k=1,2,\dots$ Given that $\lim_{t\rightarrow t_k^+}|\Delta v|=0$ and $\lim_{t\rightarrow t_{k+1}^-}|\Delta v|=\omega p^\beta$, it indicates that there exists a strictly positive constant $t^*=\frac{\omega p^\beta}{\kappa}$ such that
\begin{align}
t_{k+1}-t_k\geq t^*,~k=1,2,\dots\notag
\end{align}
with $t^*$ denoting minimal inter-transmission interval. Thereby, the Zeno behavior is excluded and the proof is completed. 
\end{proof}

\section{Coded Event-triggered Control for high-order System}

In this section, the proposed CEC will be extended to $n$th-order systems \eqref{Eq1}. Instead of step-by-step procedure, its control design is omitted here for conciseness since its stability analysis is identical to the second-order system \eqref{Eq1155}.

We first define the tracking error variables:
\begin{align}
&e_1=x_1-y_d\label{Eq57}\\
&e_i=x_i-\alpha_{i-1},~i=2,\dots,n,\label{Eq58}
\end{align}
where $\alpha_{i-1}$ represent virtual control inputs. In order to guarantee $e_1$ fulfilling the Self-adjustable Prescribed Performance \eqref{Eq5}, we introduce the following error transformation function:
\begin{align}\label{Eq59}
z_1=\ln \left( \frac{E_l+e_1}{E_u-e_1}\right)
\end{align}

Then, the virtual control inputs are designed as
\begin{align}
&\alpha_1=-\frac{1}{\mu_1}(c_1z_1+z_1\hat\theta_1\Phi_1),\label{Eq60}\\
&\alpha_i=-(c_ie_i+\hat\theta_ie_i\Phi_i),~i=2,\dots,n-1,\label{Eq61}
\end{align}
and the adaptive laws are updated by
\begin{align}
&\dot{\hat\theta}_1 = r_1z_1^2\Phi_1-\sigma_1\hat\theta_1,~~\hat\theta_1(0)\ge 0,\label{Eq62}\\
&\dot{\hat\theta}_i = r_ie_i^2\Phi_i-\sigma_i\hat\theta_i,~~\hat\theta_i(0)\ge 0,
\end{align}
for $i=2,\dots,n-1$, where $c_i$, $r_i$ and $\sigma_i$, $i=1,\dots,n-1$, are positive constants. $\hat\theta_i$ is the estimation of $\theta_i$, and $\Phi_i$ is a computable function, where $\theta_i$ and $\Phi_i$ are
\begin{align}
&\hspace{-0.2cm}\theta_i=\max \{1,b_1^2,b_2^2,\dots,b_i^2\},~i=1,\dots,n-1,\label{Eq64}\\
&\hspace{-0.2cm}\Phi_1=\mu_1^2\dot y_d^2+\mu_1^2\phi_1^2+\mu_2^2,\\
&\hspace{-0.2cm}\Phi_2=(\frac{\partial\alpha_{1}}{\partial x_1}\phi_1)^2\!+\!(\frac{\partial\alpha_{1}}{\partial x_1}x_{2})^2+(\Delta\alpha_{1})^2+\mu_1^2z_{1}^2\!+\phi_2^2,\\
&\hspace{-0.2cm}\Phi_i=\sum_{k=1}^{i-1}(\frac{\partial\alpha_{i-1}}{\partial x_k}\phi_k)^2+\sum_{k=1}^{i-1}(\frac{\partial\alpha_{i-1}}{\partial x_k}x_{k+1})^2+(\Delta\alpha_{i-1})^2\notag\\
&\hspace{-0.2cm}~~~~~~~+e_{i-1}^2+\phi_i^2,~i=3,\dots,n-1,\label{Eq67}
\end{align}
where $\mu_1=\ell(E_u+E_l)$ and $\mu_2=\ell(\dot e_l-\dot\eta_l)(E_u-e_1)-(\dot e_u+\dot\eta_u)(E_l+e_1)$ and $\ell = \frac{1}{(e_1+E_l)(E_u-e_1)}$. We can also derived that $E_u(t)+E_l(t)>0$ and $(E_u(t)-e_1(t))(E_l(t)+e_l(t))>0$ are bounded functions if $e_1(t)$ belongs to a compact set $\Omega_{e_1}=\{e_1(t)\in\mathbb{R}:-E_l(t)<e_1(t)<E_u(t)\}$. Therefore, we have $\mu_1\neq0$ and $\mu_1\in L_\infty$. Identically, the actual control input and the adaptive law are developed:
\begin{align}
&v(t)=-(c_ne_n+\hat\theta_ne_n\Phi_n),\label{Eq68}\\
&\dot{\hat\theta}_n = r_ne_n^2\Phi_n-\sigma_n\hat\theta_n,~~\hat\theta_n(0)\ge 0,\label{Eq69}
\end{align}
where $r_n,r_n,\sigma_n>0$, $\hat\theta_n$ denotes the estimation of $\theta_n$ and $\Phi_n$ is a computable function. Similar to \eqref{Eq64}-\eqref{Eq67}, we have
\begin{align}
&\theta_n=\max \{1,b_1^2,b_2^2,\dots,b_n^2\},\label{Eq70}\\
&\Phi_n=\sum_{k=1}^{n-1}(\frac{\partial\alpha_{i-1}}{\partial x_k}\phi_k)^2+\sum_{k=1}^{n-1}(\frac{\partial\alpha_{i-1}}{\partial x_k}x_{k+1})^2+(\Delta\alpha_{n-1})^2\notag\\
&~~~~~~~+e_{n-1}^2+\phi_{n}^2+1.
\end{align}
In the following theorem, we summarize the result on the Coded Event-triggered Control for the $n$th-order nonlinear systems.

\begin{theorem}
Consider $n$th-order nonlinear systems \eqref{Eq1} under Assumptions \ref{assumption1}-\ref{assumption3}. If the control inputs and the adaptive laws are designed as \eqref{Eq64}-\eqref{Eq69}, by adopting the Coded Event-triggered Scheme \eqref{Eq333}-\eqref{Eq444}, we can guarantee the following properties:
\begin{enumerate}[(i)]
\item Given any initial condition, the tracking error $e_1(t)$ is constrained by the prescribed performance: $-e_{l}(t)<e_1(t)<e_{u}(t)$, for $t\geq T$ with $T$ being a preassigned finite time, that is, $e_1$ fulfills Entry Capture Problem properties in Definition \ref{definition1};

\item All signals in the closed-loop system are guaranteed to be bounded regardless of initial conditions;

\item Zeno behavior is excluded.
\end{enumerate}
\end{theorem}

\begin{proof}
Consider the following Lyapunov candidates:
\begin{align}
&V_1=\frac{1}{2}z_1^2+\frac{\underline{g}_1}{2r_1}\tilde\theta_1^2,\notag\\
&V_i= V_{i-1}+\frac{1}{2}e_n^2+\frac{\underline{g}_n}{2r_n}\tilde\theta_n^2.\notag
\end{align}
where $\tilde\theta_i=\theta_i-\hat\theta_i,i=1,\dots,n$. Similar to the stability analysis in Theorem \ref{theorem1}, it can be also derived:
\begin{align}
\dot V_n
\leq&-c_1\underline{g}_1z_1^2-\sum_{k=2}^nc_k\underline{g}_ke_k^2-\sum_{k=1}^n\bar\sigma_k\tilde\theta_k^2+\varepsilon_n\notag\\
\leq&-\pi V_n+\varepsilon_n,\notag
\end{align}
where $\pi=\min\{2c_1\underline{g}_1,\dots,2c_n\underline{g}_n,2\frac{r_1\bar\sigma_1}{\underline{g}_1},\dots,2\frac{r_n\bar\sigma_n}{\underline{g}_n}\}>0$, $\varepsilon_n=\frac{\underline{g}_n\sigma_n}{2r_n}\theta_n^2+\frac{3}{4\underline{g}_n}+\frac{\bar g_{n-1}^2}{4\underline{g}_n}+\frac{\bar g_n^2}{4\underline{g}_n}\bar p^2+\varepsilon_1$. In accordance with \eqref{Eq333}-\eqref{Eq444}, we have $|\Delta v(t)|\leq \bar p$, where $\bar p=\max\{\omega_0 p^0,\omega_1 p,\dots,\omega_{s_c} p^{s_c} \}$ is a bounded constant. Then, we obtain
\begin{align}
0\leq V_n(t)\leq \frac{\varepsilon_n}{\pi}+(V_n(0)-\frac{\varepsilon_n}{\pi})e^{-\pi t}.\notag
\end{align}
It indicates that $V_n(t)$ is bounded, leading to $z_1$, $e_2,\dots,e_n$, $\tilde\theta_1,\dots,\tilde\theta_n\in L_\infty$. According to the error transformation function \eqref{Eq59}, we can derive that $e_1$ is evolving within the SPP, namely, $-E_l(t)<e_1(t)<E_u(t)$. Based on Lemma \ref{lemma1}, the Entry Capture Problem is achieved, that is, $-e_l(t)<e_1(t)<e_u(t)$ for all $t\geq T$.

Next, the boundedness of other signals in the closed-loop system is proved. From \eqref{Eq62}-\eqref{Eq64} and \eqref{Eq69}-\eqref{Eq70}, and the definition of $\tilde\theta_i$, it leads to $\hat\theta_i$ and $\dot {\hat\theta}_i\in L_\infty$, for $i=1,\dots,n$. By \eqref{Eq60}-\eqref{Eq61} and \eqref{Eq68}, the boundedness of control input signals $\alpha_1,\dots,\alpha_{n-1},v$ is guaranteed. Considering \eqref{Eq57}-\eqref{Eq58}, $x_1,\dots,x_n$ belong to $L_\infty$. Therefore, all signals in the closed-loop system are bounded.

To verify the exclusion of Zeno phenomenon, the time derivative of $\Delta v(t)$ before next triggering is given  
\begin{align}
 \Delta \dot v(t)=\dot v(t)-\dot v(t_k)=\dot v(t),~t_k\leq t<t_{k+1}.\notag
\end{align}
where $k=1,2,\dots$ By invoking \eqref{Eq68}, there has $\dot v(t)=\sum_{k=1}^{n}\frac{\partial v}{\partial x_k}\dot x_k+\sum_{k=0}^{n}\frac{\partial v}{\partial y_d^{(k)}}y_d^{(k+1)}+\sum_{k=1}^{n}\frac{\partial v}{\partial\hat\theta_k}\dot{\hat\theta}_k +\sum_{k=0}^{n}\frac{\partial v}{\partial \rho^{(k)}}\rho^{(k+1)}+\sum_{k=0}^n\frac{\partial v}{\partial \eta_u^{(k)}}\dot \eta_u^{(k+1)}+\sum_{k=0}^n\frac{\partial v}{\partial \eta_l^{(k)}}\dot \eta_l^{(k+1)}$. Since all the signals in the closed-loop system are bounded, from Assumptions \ref{assumption1}-\ref{assumption3}, Definition \ref{definition2}, prescribed performance functions $e_u(t)$ and $e_l(t)$ defined in \eqref{Eq8}-\eqref{Eq8}, $\dot v(t)$ is, therefore, bounded by
\begin{align}
|\Delta \dot v(t)|\leq \kappa,~t_k\leq t<t_{k+1},\notag
\end{align}
where $k=1,2,\dots$ Given that $\lim_{t\rightarrow t_k^+}|\Delta v|=0$ and $\lim_{t\rightarrow t_{k+1}^-}|\Delta v|=\omega p^\beta$, it indicates that there exists a strictly positive constant $t^*=\frac{\omega p^\beta}{\kappa}$ such that
\begin{align}
t_{k+1}-t_k\geq t^*,~k=1,2,\dots\notag
\end{align}
with $t^*$ being the minimal inter-transmission interval. Thereby, the Zeno behavior is excluded and proof is completed.
\end{proof}

\begin{remark}
\tc{We provide a guideline for the parameter selections. For the settling time $T$ in \eqref{Eq2}, its value should satisfy $0<T_{\min}<T$ where $T_{\min}$ is the minimum time necessary for signal processing. This assumption is widely used in existing results \cite{zhao2021adaptive,song2018tracking}. If $T$ is chosen too small, although faster convergence, it could lead to large control signals. Considering the control parameters $c_i$ and $\sigma_i$, their values will determine the convergence rate and tracking accuracy. Note that their extreme values may lead to large control inputs, making systems sensitive to external disturbances. Therefore, these parameters' selections should consider the balance between tracking performance and system ability.}

\end{remark}

\section{Simulation Results} \label{section6}
To illustrate the effectiveness of our proposed CEC, some comparative simulations are conducted in this section. Consider the following second-order nonlinear system:
\begin{align}\label{Eq6060}
\left\{
  \begin{array}{ll}
    \dot x_{1}=f_{1}(\bar x_{1})+g_{1}(\bar x_1)x_{2},\\
    \dot x_{2}=f_{2}(\bar x_{2})+g_{2}(\bar x_2)u,
  \end{array}
\right.
\end{align}
where $f_1(\bar x_1)=x_1^2+0.1\cos(0.5x_1)$, $f_2(\bar x_2)=4x_1x_2+x_1e^{-|x_2|}+0.05\sin(x_1x_2)$, $g_1(\bar x_1)=5 + 0.5\sin(x_1)$, and $g_2(\bar x_2)=3 + 0.2\cos(x_1x_2)$. By Assumption \ref{assumption2}, it can be derived that $|f_1(\bar x_1)|\leq x_1^2+0.1$ with $b_1=1$ and $\phi_1=1+x_1^2$, and $|f_2(\bar x_2)|\leq 3x_1^2+2x_2^2+0.3$ with $b_2=3$ and $\phi_2=x_1^2+x_2^2+1$. The initial condition is set by $x_1(0)=1.5$ and $x_2(0)=0$. The desired tracking trajectory is $y_d=\sin(0.5 t)$. The control parameters are selected as: $m=3$ (i.e., $\beta=0,1,2,3$ and $\omega =0.3,0.4,0.5,0.6$), $T=4$, $c_1=2$, $c_2=15$, $\sigma_1=\sigma_2=0.01$, $r_1=r_2=0.002$, $l=0.6$, $\lambda=2$, $\delta=0.5$, $\rho_0=1$, $\rho_\infty=0.4$, and $\varsigma=1$. Moreover, our CEC is compared with our control design with relative threshold \cite{xing2018event} and switching threshold \cite{xing2016event}.

\begin{figure}[H]
  \centering
  \includegraphics[scale=0.47]{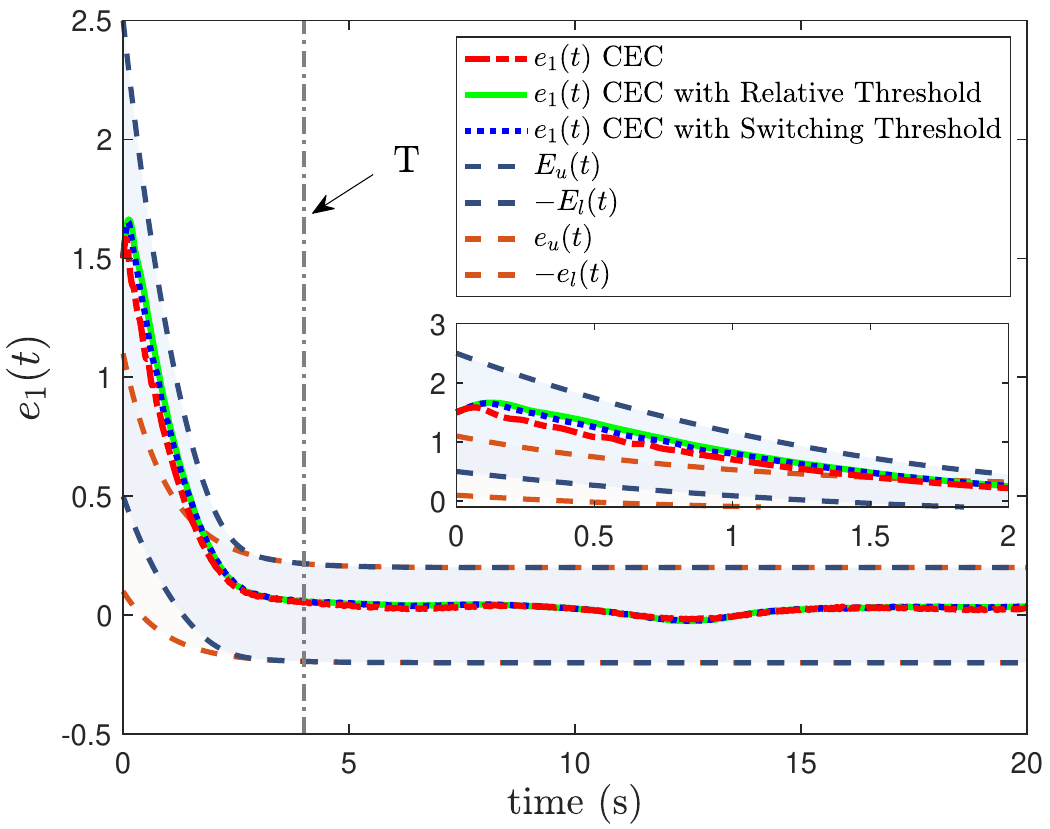}
  \caption{The tracking error $e_1(t)$ for case I.}\label{tracking_without_disturbance}
\end{figure}

\begin{figure}[H]
  \centering
  \includegraphics[scale=0.47]{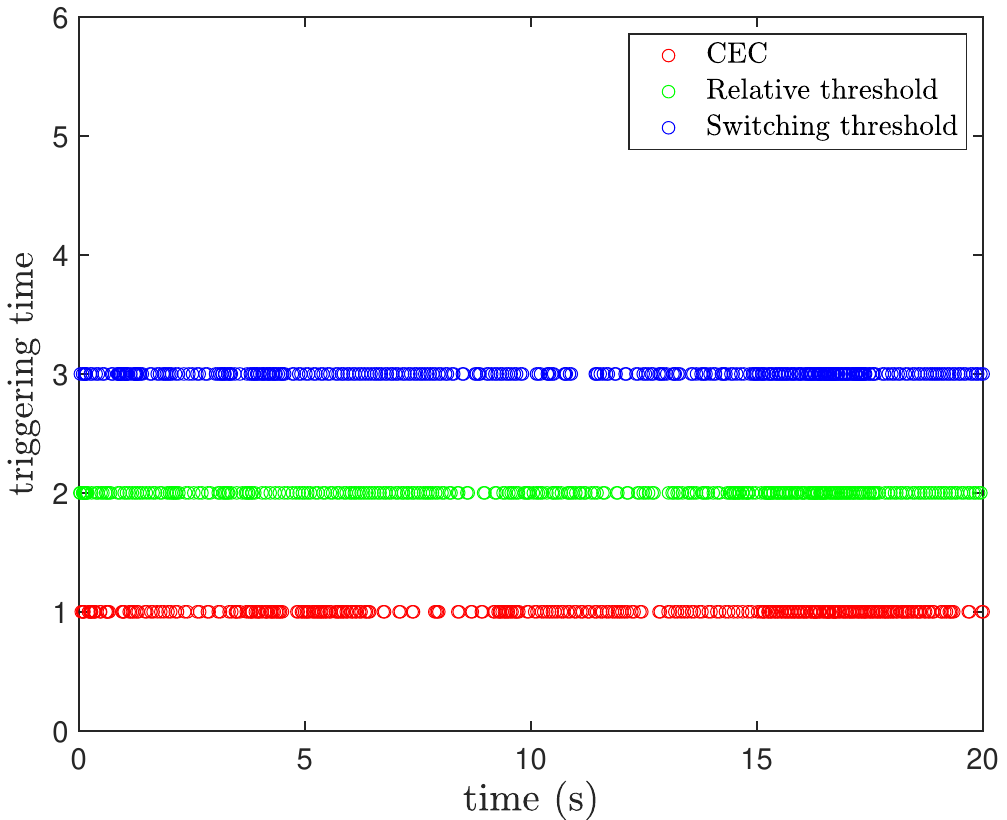}
  \caption{The triggering time for case I.}\label{triggering_time_without_disturbance}
\end{figure}

\tc{\textbf{(Case I)} The simulation results of our CEC under different threshold strategies are shown in Figs. \ref{tracking_without_disturbance}-\ref{triggering_time_without_disturbance}. From Fig. \ref{tracking_without_disturbance}, the initial tracking error $e_1(0)$ violates the initial performance constraints colored in brown, leading to the classical prescribed control methods being inapplicable. By our CEC, we can regulate $e_1(0)$ into the allowable set within the given time $T$. With the aid of SPP, bounded yet customizable virtual performance boundaries are established during the initial interval as shown in blue color. Therefore, the severe overshoot or unconstrained behaviors over $[0,T)$ are effectively avoided. On the other hand, the triggering time of three threshold strategies is shown in Fig. \ref{triggering_time_without_disturbance}, and the number of triggering events and bit assumptions are summarized in the second and third columns of Table \ref{table1}, in which we exclude the initial signal transmission. Due to $m=3$ in the CES, it indicates that only a 3-bit string is transmitted for each communication when the event condition is satisfied. Compared with the relative threshold and the switching threshold where control inputs should be encoded by 8-bit strings before each transmission, our CES consumes fewer bits while maintaining the tracking performance, which is significant for the practical system with limited communication bandwidth.
}


\begin{table}[!htbp]\label{table1}
\centering
\caption{The number of triggering events for different strategies}
\resizebox{\linewidth}{!}{
\begin{tabular}{c|c|c|c|c}
\hline
&\multicolumn{2}{|c|}{Case I}& \multicolumn{2}{|c}{Case II}\\
\hline
different strategies&trigger number&bit consumption&trigger number&bit consumption\\
\hline
CES&332&996&460&1380\\
\hline
Relative threshold&339&2712&X&X\\
\hline
Switching threshold&373&2984&403&3224\\
\hline
\end{tabular}
}
\end{table}

\begin{figure}[H]
  \centering
  \includegraphics[scale=0.47]{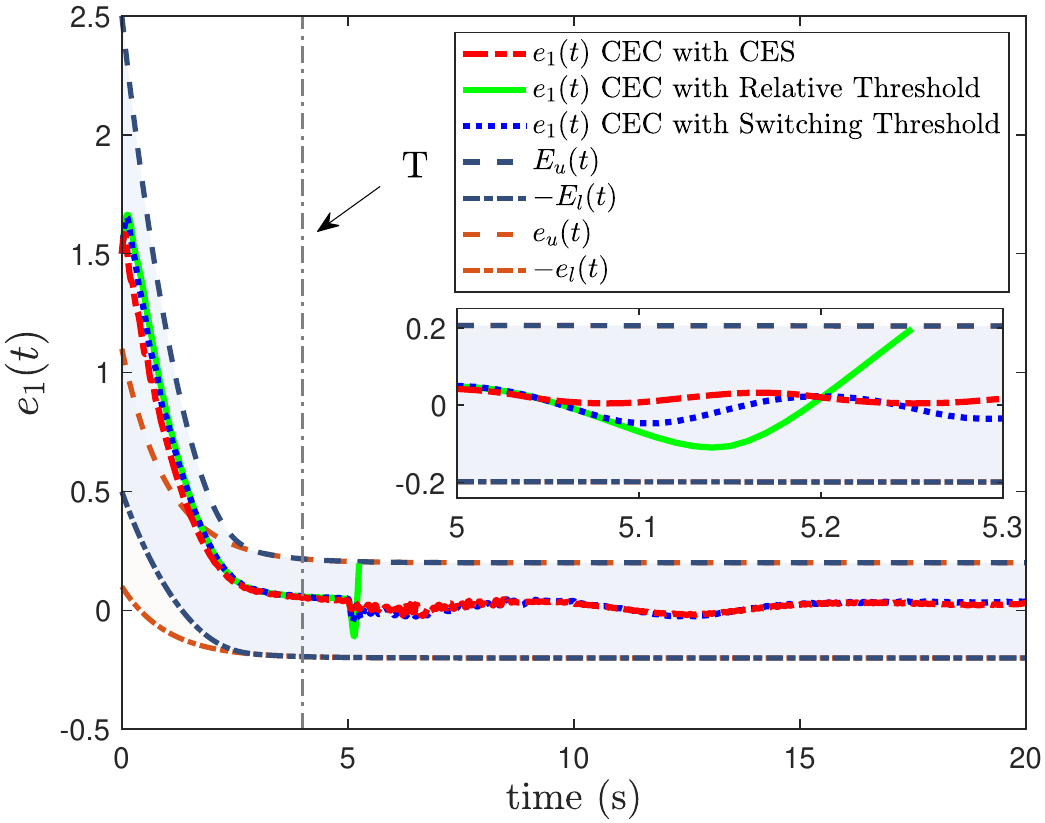}
  \caption{The tracking error $e_1(t)$ for case II.}\label{tracking_with_disturbance}
\end{figure}
\begin{figure}[H]
  \centering
  \includegraphics[scale=0.47]{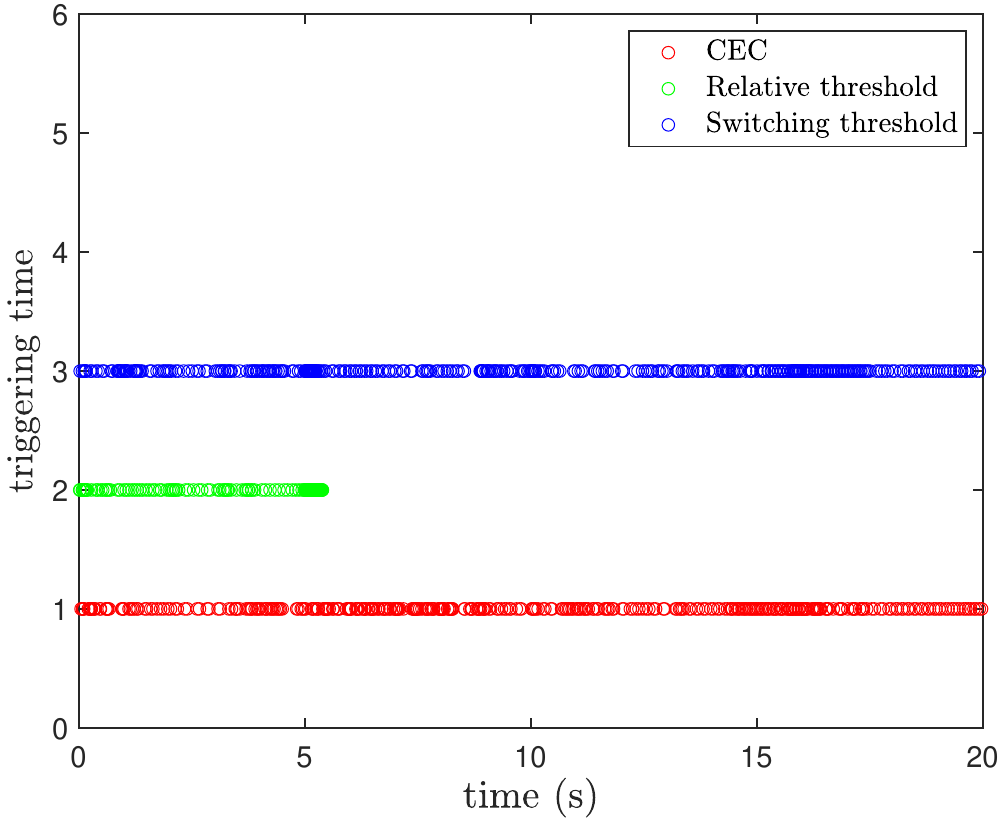}
  \caption{The triggering time for case II.}\label{triggering_time_with_disturbance}
\end{figure}

\tc{\textbf{(Case II)} To further illustrate the effectiveness of our CES, we consider the introduction of external disturbance $d(t)=2\cos0.5t$ during 5-10s in the second row of \eqref{Eq6060}. The comparative results under different threshold strategies are shown in Figs. \ref{tracking_with_disturbance}-\ref{triggering_time_with_disturbance}. Due to the signal distortion problem associated with the relative threshold strategy as discussed in Section \ref{section31}, the tracking performance is severely degraded with $e_1(t)$ depicted by the green line violating performance constraints at 5.24s, as illustrated in Fig. \ref{tracking_with_disturbance}. It makes the control design suffer from the singularity problem. Note that our CES and switching threshold strategy accomplish a good balance between tracking performance and network constraints. Therefore, they can guarantee the tracking errors within the allowable set even the introduced external disturbances. However, from the fourth and fifth columns of Table \ref{table1}  and the triggering time in Fig. \ref{triggering_time_with_disturbance}, our CES strategy can consume fewer bit resources compared to the switching threshold one. On the other hand, for each communication case, we only transmit an encrypted $3$-bit string through the public network rather than the sensitive real control input in the switching threshold strategy. Therefore, our control design reduces the communication burden and addresses the security concern at the same time.
}

\section{Conclusion}

A Coded Event-triggered Control has been designed for a class of nonlinear systems under any initial condition. We have shown that such control method can not only consume less communication bandwidth, but also enhance secure communication capability, since only $m$-length string is encoded and transmitted for each communication. An effort has been also made on developing Self-adjustable Prescribed Performance such that the initial condition-dependence restriction is removed, allowing the Entry Capture Problem to be collectively addressed. \tc{Note that our communication protocol is based on the magnitude of control input. In the future, we would like to design the protocol from the changing rate of the control signal to further enhance the tracking performance.}

\bibliographystyle{agsm}        
\bibliography{autosam}

\end{document}